\documentclass{article}[11pt]
\usepackage[a4paper, total={6in, 8in}]{geometry}
\RequirePackage{amsthm,amsmath,amsfonts,amssymb}
\RequirePackage[authoryear]{natbib}
\RequirePackage[colorlinks,citecolor=blue,urlcolor=blue]{hyperref}
\RequirePackage{graphicx}

\theoremstyle{plain}

\theoremstyle{definition}

\usepackage[utf8]{inputenc} 
\usepackage[T1]{fontenc}    
\usepackage{hyperref}       
\usepackage{url}            
\usepackage{booktabs}       
\usepackage{amsfonts}       
\usepackage{nicefrac}       
\usepackage{microtype}      
\usepackage{xcolor}         
\RequirePackage{amsthm,amsmath,amsfonts,amssymb,color,bm}

\usepackage{cleveref}

\usepackage{nameref}
\usepackage{placeins}
\usepackage{caption}
\RequirePackage{cancel}
\RequirePackage{dsfont}
\RequirePackage{mathtools}
\usepackage{algorithm}
\usepackage{algpseudocode}
\usepackage{appendix}
\usepackage{booktabs} 

\RequirePackage[english]{babel}
\RequirePackage[utf8]{inputenc}
\RequirePackage{mathrsfs}  
\RequirePackage{afterpage}
\RequirePackage{mathtools}
\RequirePackage{amsfonts}
\RequirePackage[colorinlistoftodos]{todonotes}
\RequirePackage{cleveref}
\let\cref\Cref
\RequirePackage{enumitem}
\newcommand{\excise}[1]{}

\RequirePackage{newfloat}
\usepackage{float}
\usepackage{caption}

\usepackage{subcaption}

\usepackage{adjustbox}

\usepackage{algorithm} 
\usepackage{algpseudocode} 
\usepackage{tabularx} 
\usepackage{amsmath}

\usepackage{multirow}  

\DeclarePairedDelimiterX{\infdivx}[2]{(}{)}{%
	#1\;\delimsize\|\;#2%
}

\newcommand{\RNum}[1]{\uppercase\expandafter{\romannumeral #1\relax}}
\usepackage{booktabs} 
\usepackage{caption} 

\xdefinecolor{uncblue}{rgb}{0.294,0.612,0.827}

\begin{document}

\title{Contrastive CUR: Interpretable Joint Feature and Sample Selection for Case-Control Studies}

\author{
  Eric Zhang$^1$,  Michael Love$^{1,2}$, Didong Li$^1$\\
  Department of Biostatistics University, of North Carolina at Chapel Hill$^1$\\
  Department of Genetics, University of North Carolina at Chapel Hill$^2$
}
\date{}
\maketitle

\begin{abstract}
Dimension reduction is an essential tool for analyzing high dimensional data. Most existing methods, including principal component analysis (PCA), as well as their extensions, provide principal components that are often linear combinations of features, which are often challenging to interpret. CUR decomposition, another matrix decomposition technique, is a more interpretable and efficient alternative, offers simultaneous feature and sample selection. Despite this, many biomedical studies involve two groups: a foreground (treatment or case) group and a background (control) group, where the objective is to identify features unique to or enriched in the foreground. This need for contrastive dimension reduction is not well addressed by existing CUR methods, nor by contrastive approaches rooted in PCAs. Furthermore, they fail to address a key challenge in biomedical studies: the need for selecting samples unique to the foreground. In this paper, we address this gap by proposing a Contrastive CUR (CCUR), a novel method specifically designed for case-control studies. Through extensive experiments, we demonstrate that CCUR outperforms existing techniques in isolating biologically relevant features as well as identifying sample-specific responses unique to the foreground, offering deeper insights into case-control biomedical data.
\end{abstract}

\section{Introduction}
High-dimensional data has become increasingly prevalent as advancements in technology enable the collection and analysis of large datasets. Consequently, there is a growing need for methods capable of efficiently handling and analyzing such data. Dimension reduction has emerged as a widely used technique to facilitate data analysis and visualization. One of the most established methods in this field is Principal Component Analysis (PCA,~\citealp{hotelling1933analysis}), or more generally, Singular Value Decomposition (SVD), which identifies optimal linear combinations of features to represent the data while preserving the maximum variance in a reduced space \citep{jolliffe2016principal}. PCA and its variants are commonly employed for data exploration, visualization, and downstream analyses \citep{hotelling1933analysis}. However, the linear combinations produced by PCA are often difficult to interpret, particularly in biomedical studies. For instance, the features in single-cell RNA sequencing (scRNA-seq) data correspond to genes, where a principal component (PC) obtained from PCA may be in the following form:
$$[ w_1 \cdot \text{gene$_1$} - w_2 \cdot \text{gene$_2$} + w_3 \cdot \text{gene$_3$}],$$
which provides limited practical insight for researchers seeking to identify biologically relevant genes.

For better interpretability, CUR decomposition has recently gained attention as a more interpretable dimension reduction method by selecting actual rows and columns from the data \citep{mahoney2009cur}. Unlike PCA, CUR directly identifies representative features (C for columns) and samples (R for rows) from the dataset, enhancing interpretability. Moreover, CUR offers the unique advantage of simultaneously selecting specific rows and columns, a capability lacking in traditional PCA.

However, in biomedical studies, it is often necessary to analyze two distinct groups: a foreground group (e.g., treatment or case) and a background group (e.g., control). The goal in such cases is to identify low-dimensional structures that are unique to or enriched in the foreground while minimizing shared information with the background. For example, researchers may aim to uncover disease-specific patterns while excluding features common to healthy subjects.

Several methods have been developed to address this contrastive setting. Contrastive Principal Component Analysis (CPCA) extends PCA to highlight variance specific to the foreground by removing variance in the background \citep{abid2017contrastive}. Contrastive Probabilistic Principal Component Analysis (PCPCA) combines CPCA with Probabilistic PCA (PPCA) by incorporating a probabilistic framework \citep{li2020probabilistic}. The Contrastive Latent Variable Model (CLVM) introduces a model with latent variables that can be recovered \citep{severson2019unsupervised}, while the Contrastive Poisson Latent Variable Model (CPLVM) is tailored for count data \citep{jones2022contrastive}. Despite the breadth of methods in this space, they lack the ability to simultaneously select both contrastive features and rows.

Most recently, the Contrastive Feature Selection (CFS) method has been introduced \citep{weinberger2024feature}, leveraging deep learning to achieve contrastive column selection. CFS has demonstrated superior feature selection by improving downstream prediction accuracy of subgroups within the foreground. However, CFS is limited to column selection and does not address contrastive row selection. In genetic data, Differential Expression (DE) analysis is a widely used approach for identifying genes that exhibit statistically significant differences in expression levels between groups \citep{anders2010differential, love2014moderated}. This is typically achieved through parametric tests, such as t-tests or ANOVA, or non-parametric alternatives like the Wilcoxon rank-sum test. While effective, DE analysis has notable limitations. A key drawback is its inability to account for the inherent correlation structure among genes, which can lead to an incomplete understanding of the underlying biological processes. Furthermore, DE analysis primarily focuses on identifying absolute differences in expression between groups, which may not always align with our specific interest in uncovering patterns or structures that are unique to the foreground group.

In this paper, we bridge this gap by proposing a novel method, Contrastive CUR (CCUR), which simultaneously selects features and rows that are unique to the foreground relative to the background. CCUR provides a more comprehensive view of the foreground, enabling the identification of features and samples that are distinctively relevant. We show that CCUR outperforms competing methods on one simulated dataset and three real genomic datasets in terms of both feature (genes) and sample (cell) selection. While our experiments focus on genomic data, CCUR is broadly applicable to other types of high-dimensional data.

\section{Method}

There are several proposed methods for obtaining the CUR decomposition. \citet{stewart1999four} introduced an approach based on QR factorization to derive the CUR decomposition. Another notable method involves selecting columns and rows by optimizing volume \citep{goreinov1997theory, goreinov2010find}. Additionally, the Discrete Empirical Interpolation Method (DEIM) and its variant, Q-DEIM, use partial pivoting and singular vectors to approximate CUR \citep{sorensen2016deim, drmac2016new}. This paper builds on the work of \citet{mahoney2009cur}, which also utilizes singular vectors but offers a more intuitive interpretation. Below, we briefly introduce this algorithm.

The SVD of a general matrix $X \in \mathbb{R}^{n \times p}$ is given by 
$X = U \Sigma V^T$, where $U \in \mathbb{R}^{n \times n}$, $\Sigma = \operatorname{diag}(\sigma_1, \dots, \sigma_\rho) \in \mathbb{R}^{n \times p}$, and $V = [v^1, \dots, v^p]$ with $\{v^t\}_{t=1}^p \in \mathbb{R}^p$. The columns of $V$ represent the right singular vectors. For each column $d$ of $X$, we define the leverage score as:
\begin{equation}
    \label{eq:leverage}
    l_d = \sum_{\xi=1}^k (v_d^\xi)^2,
\end{equation}
where $k$ denotes the top $k$ right singular vectors. Since the columns of $V$ are orthonormal, the leverage scores correspond to the diagonal elements of the projection matrix onto the subspace spanned by the top $k$ right singular vectors. Thus, each score quantifies the influence or leverage that a column exerts on the optimal low-rank approximation of the matrix. By calculating these leverage scores, we can prioritize columns that have a disproportionately large impact on the low-rank structure of $X$.

To select rows, the same procedure can be applied to the transpose of $X$. Once the matrices $C$ and $R$ are constructed, we carefully compute $U$ such that the product $CUR$ closely approximates $X$. The CUR algorithm is detailed in \Cref{apdx:alg}.
 CUR can be adapted to randomly sample columns and rows by normalizing the leverage scores to create a probability distribution. This approach utilizes oversampling and randomness, providing strong guarantees for matrix approximation \citep{drineas2006fast}.

However, in this paper, our focus is on identifying salient features and samples specific to the foreground group. As such, we do not emphasize optimal matrix decomposition. Instead, in our proposed CCUR method, we use fixed parameters $c$ and $r$ to deterministically select a predefined number of columns and rows, offering users precise control over the selection process.

Building on the concept of leverage scores, we extend CUR to the contrastive setting, where data is divided into two groups: foreground and background. Define $\{x_i\}_{i=1}^n, \{y_i\}_{i=1}^m$ where $x_i, y_j \in \mathbb{R}^p$ as the foreground and background data respectively. We aim to prioritize columns that are especially salient in the foreground while being uninformative in the background.

CCUR calculates the standard leverage scores, defined in \cref{eq:leverage}, for both the foreground and background, using the top $k$ right singular vectors. Next, for each column $d \in \{1, \dots, p\}$, we compute a contrastive score:
\begin{equation*}
    l_d = \frac{l^x_{d}}{l^y_{d} + \epsilon},
\end{equation*}
where $l^x_d$ and $l^y_d$ are the leverage scores in the foreground and background groups, respectively, and $\epsilon > 0$ is a small constant added for numerical stability. The top $c$ columns with the highest contrastive scores and then selected. This selection favors features that are more important in the foreground (high leverage) but negligible in the background.

To prevent inflated ratios due to near-zero background scores, we introduce a small constant $\epsilon$ to the denominator. Specifically, when $l^y_d \approx 0$, the resulting ratio becomes disproportionately large, even if the corresponding foreground leverage score is also low. This can lead to the erroneous selection of features that are not truly salient in the foreground. By introducing $\epsilon$, we mitigate this issue, ensuring that features with negligible leverage in both groups are appropriately deprioritized. The CCUR column selection algorithm is detailed in \cref{alg:ccur_col}.

\begin{algorithm}[ht]
\caption{Contrastive Column Selection}
\begin{algorithmic}[1]
\State \textbf{Input} foreground and background data: $\{x_i\}_{i=1}^n$, $\{y_j\}_{j=1}^m$, number of singular vectors $k$, number of features selected $c$, positive constant $\epsilon$.
\State Compute the SVD of $X$ and $Y$.
\State Compute leverage scores of each column in the target and background.
$$l^x_d = \sum_{\xi = 1}^k (v_{x}^{\xi})_d^2 , \quad d = 1 \ldots p,
$$
$$l^y_d = \sum_{\xi = 1}^k (v_{y}^{\xi})_d^2 , \quad d = 1 \ldots p,
$$
where $v_{x}^{\xi}, v_{y}^{\xi}$ are the $\xi$th right singular vectors the  of the foreground and background respectively.
\State Compute ratio of leverage scores.
$$l_d = \frac{l^x_{d}}{l^y_{d} + \epsilon}, \quad d = 1  \ldots p.$$

\State Rank the ratios and return the indices of the top $c$ ratios. 
\end{algorithmic}
\label{alg:ccur_col}
\end{algorithm}

To perform contrastive row selection, we restrict the dataset to the $c$ columns selected via contrastive column ranking. CUR is then applied to this reduced matrix and we select rows based on its leverage scores. By narrowing the focus to a smaller subset of distinctive features that are highly relevant in the foreground and minimally relevant in the background, CUR can effectively capture rows that are uniquely associated with the foreground group. This is summarized in \cref{alg:ccur_row}.

\begin{algorithm}[!h]
\caption{Contrastive Row Selection}
\begin{algorithmic}[1]
\State \textbf{Input} target $\{X_i\}_{i=1}^n$, $C$: the set of indices returned in \cref{alg:ccur_col}, number of singular vectors $k$, number of rows selected $r$.
\State Run CUR on $X_{C}$ and return R.
\end{algorithmic}
\label{alg:ccur_row}
\end{algorithm}

\section{Simulation}
In this section, we present a simulation setting to showcase CCUR's ability to recover foreground-specific information through column and row selection. Let $\{x_i\}_{i=1}^n \in \mathbb{R}^{p}$ and $\{y_j\}_{j=1}^m \in \mathbb{R}^{p}$
be the foreground and background data respectively. Following a similar philosophy in CLVM, we generate the data as follows.

\begin{figure}[!h]
    \centering
  \includegraphics[width=0.8\linewidth]{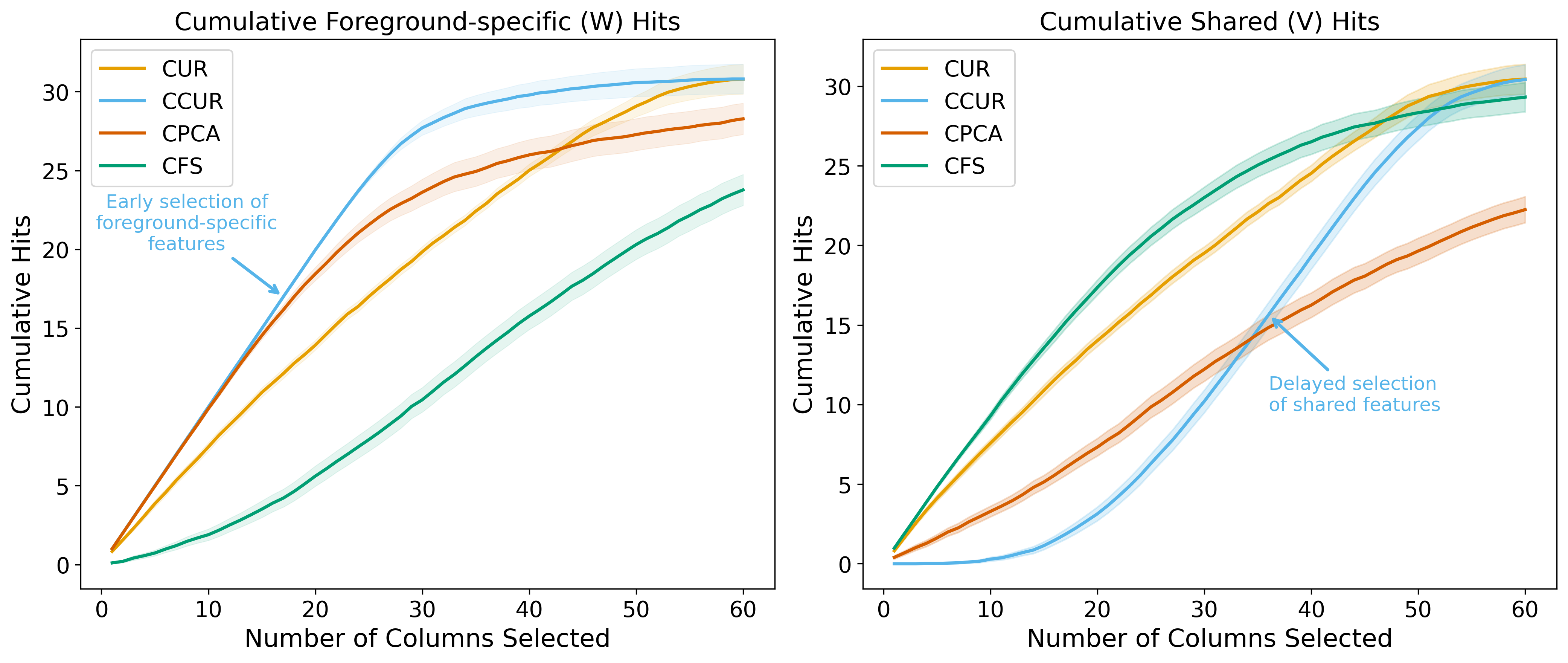}
  \includegraphics[width=0.8\linewidth]{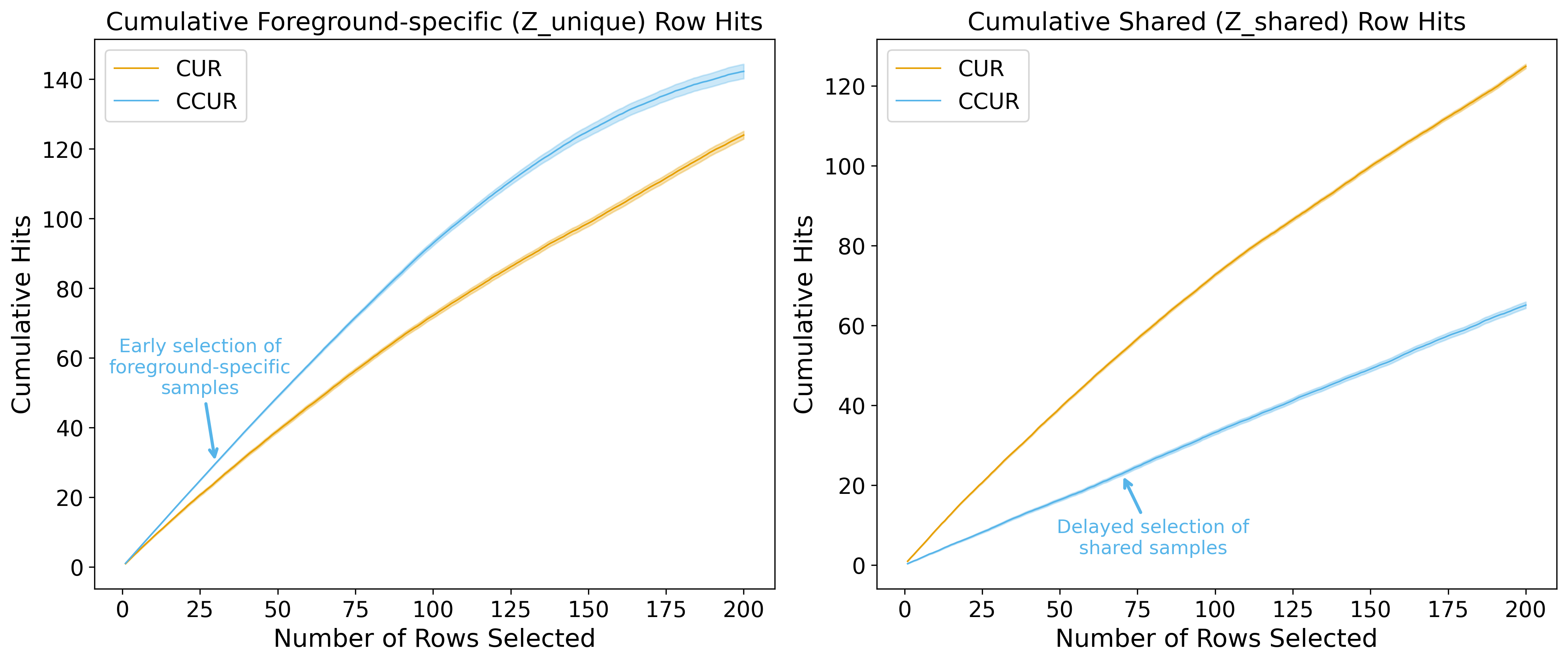}
     \caption{Cumulative recovery of true signal  evaluated on simulated data. Top left: the number of foreground-specific ($W$-associated) columns recovered among the top-ranked columns. Top right: recovery of shared ($V$-associated) columns. Bottom left: number of foreground-specific ($Z_{\text{unique}}$-associated) rows recovered among the top-ranked rows. Bottom right: recovery of shared ($Z_{\text{shared}}$-associated) rows. Shaded regions represent $\pm 2$ standard errors computed across 100 independent simulation replicates.}
     \label{fig:sim}
\end{figure}

\begin{align}
\label{eq:sample}
x_i &= Vz_i + Wt_i +\epsilon_i, \quad i = 1, \dots, n,\nonumber\\
y_j &= Vs_j +\epsilon_j, \quad j = 1, \dots, m,\nonumber
\end{align}
where $z_i, t_i, s_j \in \mathbb{R}^k \sim \mathcal{N}(0,I_k)$ represent the latent variables, and $\epsilon_i, \epsilon_j \in \mathbb{R}^p \sim \mathcal{N}(0,I_p)$ are the noise. $V, W \in \mathbb{R}^{p \times k}$ also have entries independently drawn from $\mathcal{N}(0, 1)$, which represent the loading matrices.
Let $Z_{\text{unique}} \in \mathbb{R}^{n \times k}$ and $Z_{\text{shared}} \in \mathbb{R}^{n \times k}$ denote the matrices whose $i$th rows are given by $t_i^\top$ and $z_i^\top$ respectively. We induce sparsity in \( V \), \( W \), \( Z_{\text{fg}} \), and \( Z_{\text{shared}} \) by applying elementwise thresholding, setting each entry \( x \) to zero if \( |x| < 1.8 \).

 For column selection, we run CCUR and compare its performance to three baseline methods: CUR, CPCA, and CFS. We evaluate each method based on the cumulative recovery of rows with at least one nonzero entry in the $W$ and $V$ matrices, corresponding to foreground-specific and shared components, respectively (\cref{fig:sim} top row). CCUR consistently recovers nonzero $W$ rows at a faster rate than CUR and CFS, while simultaneously accumulating fewer nonzero $V$ rows. This demonstrates CCUR’s effectiveness in isolating foreground-specific features. Although CPCA performs competitively, it relies on a heuristic that selects features with the highest absolute loadings in the first principal component, which may be suboptimal in settings where contrastive variation is not aligned with a single dominant direction. A similar trend is observed in row selection: CCUR more efficiently identifies rows with at least one non-zero entry in $Z_{\text{unique}}$, whereas CUR tends to favor non-zero in $Z_{\text{shared}}$ The results are displayed in the bottom row of \cref{fig:sim}. These results highlight the ability of CCUR to identify the unique structures in both feature and sample dimensions.





\section{Applications}

We apply our method to three real-world biomedical datasets. Through these applications, we demonstrate that CCUR not only identifies subtle, foreground-enriched genes/proteins but also preferentially selects representative samples. We primarily focus on scRNA-seq data where features correspond to genes and samples correspond to cells.

A key consideration in applying CCUR is the selection of foreground and background datasets. Generally, the background dataset should contain information that we aim to filter out from the foreground, thereby isolating the unique features of the foreground. In case-control studies, it is natural to designate the case group as the foreground, as the goal is to capture variations specific to the case group.

Several parameters in CCUR require careful consideration. For instance, the number of singular vectors, $k$, plays a crucial role. Additionally, the number of genes to be selected must be specified. In our experiments, we select $c = 10$ genes, although this number can be adjusted as needed. To benchmark CCUR, we compare its performance against CUR applied to the foreground and the union of both groups, as well as other contrastive methods like CPCA and CFS. For additional discussion on experimental detail and hyperparameter selection, see \cref{appendix:hyper} and \cref{appendix:experiments}.

\subsection{Mouse Protein}
\label{sec:mp}
\begin{figure}[h!]
    \centering
\includegraphics[width=\linewidth]{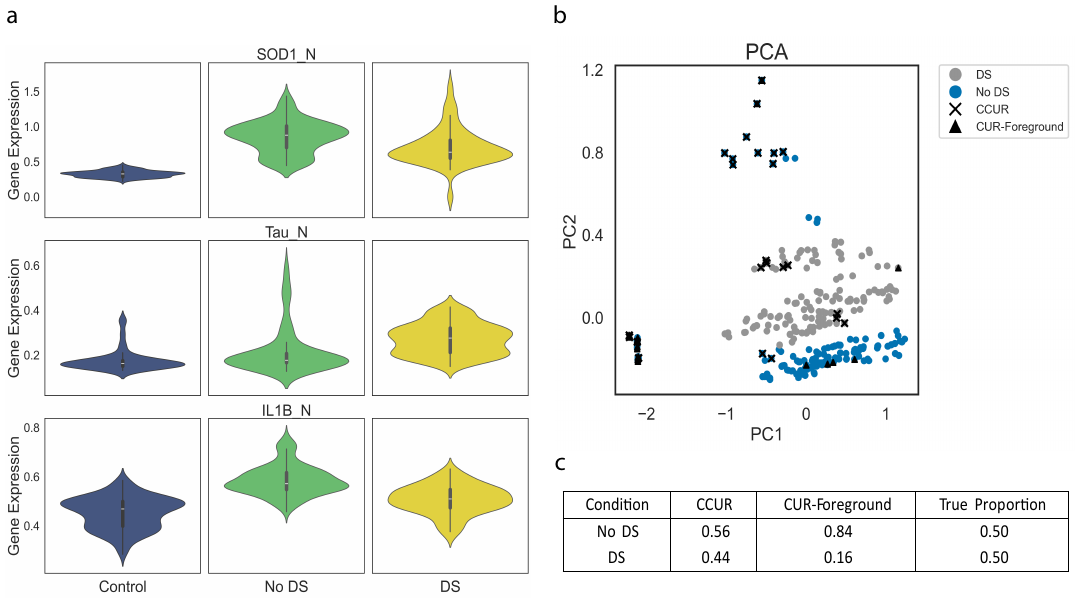}
    \caption{(a) Gene expressions for \textit{Sod1}, \textit{Tau}, and \textit{Il1b} across the control, no DS, and DS groups. (b) Top 25 rows selected. PCA performed on the 10 selected columns from contrastive column selection. (c) Proportion of rows selected. Comparison of row selection from CCUR, CUR applied to Foreground, and the true proportion of DS and no DS mice.}
    \label{fig:mouse_protein}
\end{figure}

The first dataset we analyze consists of protein expression data from mice \citep{higuera2015self}. The foreground group includes mice, both with and without Down Syndrome (DS), that underwent shock therapy to stimulate learning, while the background group is the control group which consists of mice without DS that did not receive shock therapy. Our goal is to identify not only genes that are unique to the foreground but also representative samples of mice within this group. To illustrate how the genes selected by CCUR can yield new biological insights, we visualize the expression patterns of several genes identified by CCUR in \cref{fig:mouse_protein} and perform relevant biological analyses. We display the control or background and the two subgroups within the foreground. These analyses also compare the genes selected by CCUR to those identified by other methods. Additional details can be found in \cref{appendix:mp}.

Among the top genes, \textit{AcetylH3K9} and \textit{Bcatenin} are the overlapping genes selected by CCUR and CUR, and we found both have shown to play an important role in learning and memory \citep{mills2014cognitive, peixoto2013role}. Overall, we found that genes selected by CUR and CPCA tend to have functions related to general cognitive functions such as \textit{pCamkii, Nr2a, Psd95} \citep{yasuda2022camkii, mcquail2016nr2a, coley2018psd95}. \textit{Pkcg}, selected by CUR, CPCA, and CFS, has been shown to be involved in neuropathic pain development \citep{ncbi_prkcg}.

Next, we consider several genes selected only by CPCA. \textit{Gsk3b} regulates dopamine receptor signaling, which controls several functions such as cognition and emotion \citep{yu2023role}. \textit{pNr2b} has been shown to regulate learning, memory, and pain perception \citep{loftis2003n}. Variation between the foreground and background however is minimal.

Then, we analyze several genes selected by CFS. \textit{Dyrk1a} plays a role in brain development and also is related to DS \citep{atas2021dyrk1a}. Analyzing the change in expression for \textit{Dyrk1a}, there is little difference among the two subgroups in the foreground but there is a clear difference between the background and foreground. \textit{Mek} has been shown to play important roles in learning and memory \citep{shalin2004neuronal}, while \textit{Trka} is a nerve growth factor that is critical for the developing of neurons including pain-mediating neurons \citep{luberg2015novel}, which could be relevant to the foreground but we observe very little change among the groups and subgroups.

The genes selected by CCUR only, however, have clear associations with shock therapy, fear conditioning, and DS. For instance, \textit{Sod1} is involved with negative inflammatory responses and oxidate stress and is crucial in neural dysfunction in DS  \citep{iannello1999oxidative}. \textit{Tau} is shown to exacerbate stress responses and negatively regulate neuron death, including environments that induce physical or psychological stress \citep{liu2023tau}. Moreover, \textit{Il1b} also plays a similar role in the central nervous system by inducing various behavioral changes and responses to stress \citep{yin2024interleukin}. 

Upon further inspection of the selected genes by CCUR, we observe that not only is there a clear difference in expression between the foreground and background, but also between the DS and no DS mice within the foreground. Each of the genes displayed in Fig. 1a has different distributions in expression across each group. This crucial finding highlights the effect of shock therapy and more importantly, its effect in the presence of DS. These findings underscore CCUR’s capability to extract nuanced patterns within the foreground. 

To analyze salient rows (mice) in the foreground group, we compare the rows selected by CCUR and those from CUR on the foreground data as the baseline. To visualize the selected mice, we project the contrastive gene selections onto the first two principle components of PCA, and highlight the mice selected by CCUR and CUR. The results are summarized in Fig 1b and 1c. Most notably, we observed CUR tends to over select mice without DS, while CCUR selects relatively more mice from the DS group. This is critical for downstream analysis since one of the objectives is to study how mice with DS perform cognitive tasks in the context of shock therapy. CCUR on the other hand selects relatively more mice that have DS, enabling us to do meaningful downstream analysis since the resulting samples are more representative of the foreground.

\subsection{Small Molecules}
\label{sec:sm}

\begin{figure}[h!]
    \centering
\includegraphics[width=\linewidth]{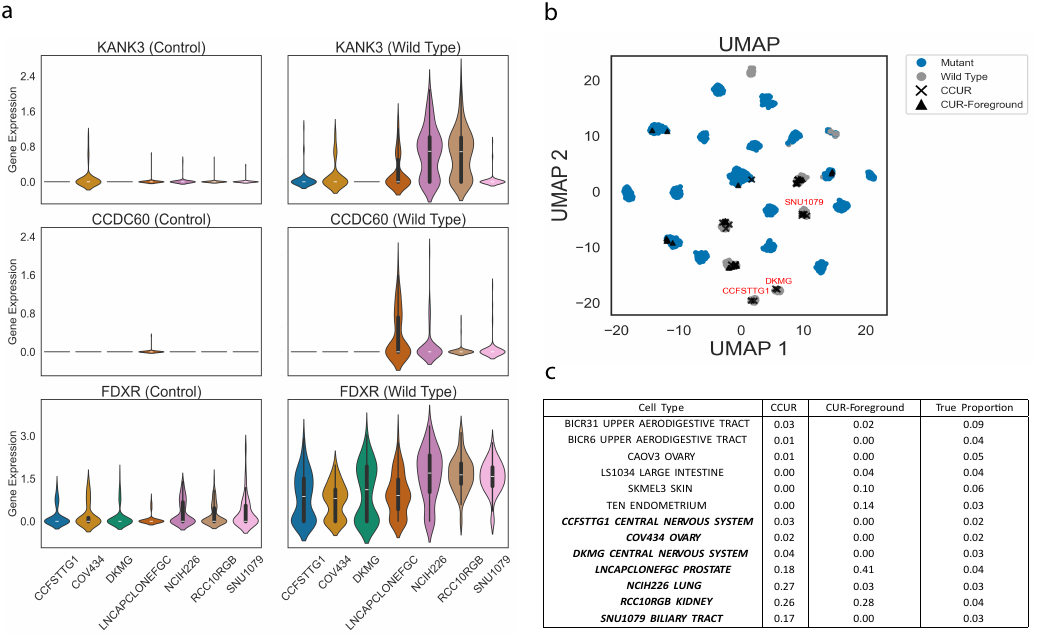}
    \caption{(a) Gene expressions for \textit{Kank3}, \textit{Ccdc60}, and \textit{Fdxr} across the control, and WT Cells. (b) UMAP applied to foreground, grouped by cell type. CCUR selected rows and CUR selected rows are shown. For clarity, only a maximum of five selected cells for a given type is shown. The 3 WT cells labeled in red are WT cells selected by CCUR but not CUR. (c) Proportion of cell types selected at least once by each method. Comparison of CCUR, CUR applied to Foreground, and the true proportion. WT cells are bolded and italicized. 200 rows were selected.}
    \label{fig:sm}
\end{figure}

The second dataset analyzed was collected using the MIX-Seq platform \citep{mcfarland2020multiplexed}, which captures transcriptional responses across more than 100 cancer cell lines. This study focuses on 24 cell lines, where the foreground group consists of cells treated with idasanutlin, a small molecule antagonist of \textit{Mdm2}, a key negative regulator of the tumor suppressor protein p53. Idasanutlin presents significant therapeutic potential in cancer by specifically activating the p53 pathway in cell lines with functional, Wild Type (WT) TP53, while this activation is absent in cell lines with transcriptionally inactive mutant TP53 \citep{vassilev2004vivo}. The background group serves as a control, comprising the same cell lines treated with dimethyl sulfoxide (DMSO). Our objective is to identify genes involved in the p53 pathway and to select cell types that capture the unique transcriptional characteristics of the foreground group. Several genes selected by CCUR are displayed in \cref{fig:sm}. For additional analysis, see \cref{sec:pathogen}.

Since activation of the p53 pathway was only observed in WT cells, we only show gene expressions in WT cells here for clarity. We found that several of the genes selected by only CUR are fundamental to cellular function and their expression is often related to general cell growth and division, and therefore not specific to the foreground (\textit{Rpl19}, \textit{Rps4x}, \textit{Lgals9}).

For the genes selected by CFS, several showed upregulation in specific WT cells. For instance, \textit{S100a2}, a tumor suppressor involved in various cellular processes, including cell cycle progression \citep{sugino2022influence}, was notably upregulated in RCC10RGB (Lung) and SNU1079 (Biliary Tract) WT cells. \textit{Tgm2}, a key effector in the TP53 tumor-suppressive pathway through its role in inducing autophagy \citep{yeo2016transglutaminase}, exhibited upregulation specifically in RCC10RGB cells. Furthermore, \textit{H4c3}, selected by both CPCA and CFS, is an essential protein for nucleosome structure \citep{ncbi_h4c3}. Interestingly, its expression was consistently downregulated across all WT cells.

In contrast, the genes selected by CCUR stand out as key markers of significant change in WT cells. For instance, \textit{Fdxr}, a direct target of the p53 family activated during DNA damage via specific p53 binding sites \citep{liu2002ferredoxin}, is consistently upregulated across all WT cells. We also observe that CCUR identifies genes that are highly sparse in the background, such as \textit{Kank3} and \textit{Ccdc60} (Fig. 2a). \textit{Kank3} mediates the p38 MAPK pathway, which interacts with p53 to regulate stress response pathways \citep{roy2018role, dai2023kank3}, while \textit{Ccdc60} has shown associations with tumor immunity, indicating its regulation is dependent on  p53 \citep{liu2023comprehensive}. These findings highlight the central role of p53 in modulating the expression of these genes, which are implicated in essential processes such as DNA damage response, apoptosis, and cell cycle regulation. The results demonstrate CCUR's strength in identifying biologically significant genes within the p53 signaling framework. We note that CPCA is able to select genes that overlap with CCUR (e.g. \textit{Fdxr}, \textit{Sugct}) but it fails to identify genes that are specifically sparse in the background such as \textit{Ccdc60}. 

Next, we analyze the contrastive rows selected by CCUR. The results are summarized in Fig. 2b and 2c. For clarity, we only display the cell types that we selected by least once by either method. 

CCUR selects each of the seven WT cells, reinforcing the finding that idasanutlin specifically activates the p53 pathway in these cells. Notably, CCUR emphasizes \textbf{NCIH226} (Lung), \textbf{RCC10RGB} (Kidney), and \textbf{SNU1079} (Biliary Tract) cell types, aligning with the observation that several genes selected by CCUR exhibit significant regulation in these lines. This pattern, while not explicitly highlighted in \citep{mcfarland2020multiplexed}, emphasizes CCUR's capability to identify cells uniquely representative of the foreground. Moreover, CCUR is able to capture cell-specific responses, moving beyond merely comparing overall differences between the foreground and background. In contrast, while CUR captures some WT cells, it fails to select the complete set and selects cells like Ten Endometrium and Skmel3 Skin, which lack specificity to the foreground. 

\subsection{Pathogen}
\label{sec:pathogen}
The final dataset we analyzed examines gene expression in the epithelial cells of mice infected with Salmonella enterica (Salmonella) or Heligmosomoides polygyrus (H. poly), as studied in \cite{haber2017single}. The background consists of healthy cells. Our objective is to identify genes that capture unique information associated with these infections. Several genes selected by CCUR are presented in \cref{fig:pathogen}. For further experimental details, see \cref{appendix:pathogen}.

\begin{figure}[h!]
    \centering
\includegraphics[width=\linewidth]{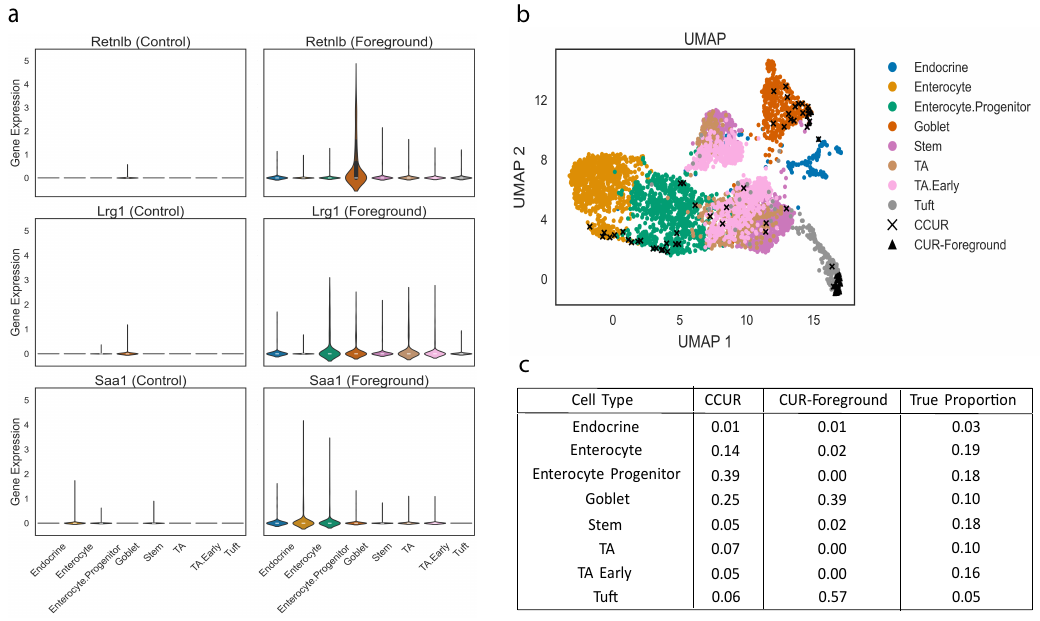}
    \caption{(a) Gene expressions compared between control and foreground for (a) \textit{Retnlb} (b) \textit{Lrg1}, (c) \textit{Saa1}. (b) UMAP applied to foreground, grouped by cell type. CCUR selected rows and CUR selected rows are shown. For clarity, top 50 rows are displayed. (c) Proportion of cells selected at least once by each method. Comparison of CCUR, CUR applied to Foreground, and the true proportion. 200 rows were selected.}
    \label{fig:pathogen}
\end{figure}

\textit{Reg3g}, selected by both CUR and CPCA, has been previously identified as a protective gene during infections like Salmonella \citep{loonen2014reg3gamma} and is known to be upregulated in enterocytes. Additionally, \textit{Apoa1}, \textit{Fabp1}, \textit{Defa24}, and \textit{Guca2b}, also selected by CUR and CPCA, were shown to be upregulated in Salmonella-infected cells compared to H. poly cells \citep{haber2017single}.

Among the top genes identified by CFS were \textit{Cd74} and \textit{Ang}, both critical regulators of the innate immune response \citep{logunova2023h2, hooper2003angiogenins}. \textit{S100a6}, another gene selected by CFS, plays a key role in maintaining the structural framework of cells and supporting cell growth \citep{wang2023s100a6}. Further analysis revealed that some of these genes show modest upregulation in certain cells within the foreground compared to the background.

Notably, for genes uniquely selected by CCUR, nearly all exhibited sparse or near-zero expression in the background but showed notable changes in expression distribution in the foreground. For instance, as shown in Fig. 3a, \textit{Retnlb} was sparse across almost all cells in the background but displayed small yet distinct expression changes in the foreground, especially in Goblet cells. \textit{Retnlb} is closely associated with immune responses, particularly in the expulsion of gastrointestinal parasites like H. poly \citep{shi2023resistin}. Similar patterns were observed for \textit{Saa1} and \textit{Lrg1}, both of which are key players in innate immunity and disease pathogenesis, including Salmonella infections \citep{hari2005serum, camilli2022lrg1}. Additionally, \textit{Saa1} was observed to exhibit enterocyte-specific expression in response to Salmonella infection, as highlighted in \cite{haber2017single}.

CCUR tends to select genes that are highly sparse in the background, genes that are often disregarded or filtered out in standard analyses, but exhibit subtle yet meaningful distributional changes in the foreground. These low-level activation patterns, though potentially rare, may carry significant biological insights that are often overlooked by conventional methods. In the following analysis, we demonstrate that these selected genes can be leveraged to identify cell types that are most representative of the foreground.

To visualize the selected cells, we project the data onto two dimensions using UMAP (Fig. 3b). The distribution of selected cells across different cell types is summarized in Fig. 3c. Notably, CCUR demonstrates a superior ability to select a more diverse set of cells compared to CUR. Importantly, CCUR emphasizes the selection of Enterocyte, Enterocyte Progenitor, and Goblet cells, cell types that we observed to exhibit the most significant changes in expression between the background and foreground, underscoring its ability to detect cell-specific changes in the foreground. In contrast, CUR heavily favors Tuft cells while neglecting Enterocyte Progenitor cells altogether, resulting in a less comprehensive selection of cell types. Although CUR does select a substantial number of Goblet cells, it overlooks the nuanced changes in other critical cell populations. 

Moreover, further analysis revealed that these cell types have well-established associations with infections. Enterocytes, and similarly Enterocytes Progenitor, are essential not only for digestive functions but also for preventing bacteria from penetrating deeper tissue layers and epithelial repair, serving as a critical line of defense  against pathogens \citep{snoeck2005role, rees2020regenerative}. 
Similarly, goblet cells contribute to immune defense by producing a protective mucus layer that shields against pathogen invasion \citep{dao2023gobletcells}. TA and TA Early cells, which were selected by CCUR and not CUR, play a vital role in maintaining intestinal integrity, acting as a rapid-response system for epithelial regeneration, helping to replenish lost or damaged cells \citep{hsu2014transit}.

\section{Discussion}

In this paper, we introduced Contrastive CUR (CCUR), a novel method designed for case-control studies to simultaneously select salient features and samples in the foreground relative to the background. By implementing contrastive leverage scores, CCUR provides a more interpretable and biologically relevant dimension reduction compared to traditional methods such as CPCA. Through comprehensive simulations and real-world experiments on scRNA-seq data, we demonstrated CCUR’s ability to isolate genes and cells that are distinct to the foreground group, which offers new insights into biological processes and disease mechanisms. In particular, we observed CCUR's ability to select genes that are sparse in the background, which are often overlooked in traditional analyses, and cells that exhibit subtle yet distinct expression changes in specific cells within the foreground. This highlights CCUR's capacity to uncover biologically meaningful variations that may otherwise go unnoticed, offering deeper insights into cell-specific gene activity and its potential role in the underlying biological processes. 

To enhance the utility and applicability of CCUR, several promising directions can be pursued. Extending CCUR to
manage scenarios with multiple foreground groups, particularly in biomedical studies where multiple treatment
groups are compared against a single control. In addition, developing methods for automated tuning of hyperparameters, such as the number of singular vectors ($k$) and the up-weighting constant ($\epsilon$), would simplify the process for users who may not be experts in statistical methodology. 

\bibliographystyle{chicago}
\bibliography{ref}

\begin{thebibliography}{}

\bibitem[\protect\citeauthoryear{Abid, Zhang, Bagaria, and Zou}{Abid et~al.}{2017}]{abid2017contrastive}
Abid, A., M.~J. Zhang, V.~K. Bagaria, and J.~Zou (2017).
\newblock Contrastive principal component analysis.
\newblock {\em arXiv preprint arXiv:1709.06716\/}.

\bibitem[\protect\citeauthoryear{Anders and Huber}{Anders and Huber}{2010}]{anders2010differential}
Anders, S. and W.~Huber (2010).
\newblock Differential expression analysis for sequence count data.
\newblock {\em Nature Precedings\/}, 1--1.

\bibitem[\protect\citeauthoryear{Atas-Ozcan, Brault, Duchon, and Herault}{Atas-Ozcan et~al.}{2021}]{atas2021dyrk1a}
Atas-Ozcan, H., V.~Brault, A.~Duchon, and Y.~Herault (2021).
\newblock Dyrk1a from gene function in development and physiology to dosage correction across life span in down syndrome.
\newblock {\em Genes\/}~{\em 12\/}(11), 1833.

\bibitem[\protect\citeauthoryear{Camilli, Hoeh, De~Rossi, Moss, and Greenwood}{Camilli et~al.}{2022}]{camilli2022lrg1}
Camilli, C., A.~E. Hoeh, G.~De~Rossi, S.~E. Moss, and J.~Greenwood (2022).
\newblock Lrg1: an emerging player in disease pathogenesis.
\newblock {\em Journal of biomedical science\/}~{\em 29\/}(1), 6.

\bibitem[\protect\citeauthoryear{Chen, Yang, Hwang, and Yang}{Chen et~al.}{2018}]{chen2018emerging}
Chen, S.-D., J.-L. Yang, W.-C. Hwang, and D.-I. Yang (2018).
\newblock Emerging roles of sonic hedgehog in adult neurological diseases: neurogenesis and beyond.
\newblock {\em International journal of molecular sciences\/}~{\em 19\/}(8), 2423.

\bibitem[\protect\citeauthoryear{Coley and Gao}{Coley and Gao}{2018}]{coley2018psd95}
Coley, A.~A. and W.-J. Gao (2018).
\newblock Psd95: A synaptic protein implicated in schizophrenia or autism?
\newblock {\em Progress in Neuro-Psychopharmacology and Biological Psychiatry\/}~{\em 82}, 187--194.

\bibitem[\protect\citeauthoryear{Dai, Xie, Yang, Chen, Hu, and Chen}{Dai et~al.}{2023}]{dai2023kank3}
Dai, Z., B.~Xie, B.~Yang, X.~Chen, C.~Hu, and Q.~Chen (2023).
\newblock Kank3 mediates the p38 mapk pathway to regulate the proliferation and invasion of lung adenocarcinoma cells.
\newblock {\em Tissue and Cell\/}~{\em 80}, 101974.

\bibitem[\protect\citeauthoryear{Dao and Le}{Dao and Le}{2023}]{dao2023gobletcells}
Dao, D. and P.~Le (2023).
\newblock Histology, goblet cells.
\newblock Updated 2023 Mar 15. In: StatPearls [Internet]. Treasure Island (FL): StatPearls Publishing; 2024 Jan-.

\bibitem[\protect\citeauthoryear{Deng, Matsuda, Tanikawa, Lin, Furukawa, Hamamoto, and Nakamura}{Deng et~al.}{2014}]{deng2014late}
Deng, Z., K.~Matsuda, C.~Tanikawa, J.~Lin, Y.~Furukawa, R.~Hamamoto, and Y.~Nakamura (2014).
\newblock Late cornified envelope group i, a novel target of p53, regulates prmt5 activity.
\newblock {\em Neoplasia\/}~{\em 16\/}(8), 656--664.

\bibitem[\protect\citeauthoryear{Drineas, Kannan, and Mahoney}{Drineas et~al.}{2006}]{drineas2006fast}
Drineas, P., R.~Kannan, and M.~W. Mahoney (2006).
\newblock Fast monte carlo algorithms for matrices iii: Computing a compressed approximate matrix decomposition.
\newblock {\em SIAM Journal on Computing\/}~{\em 36\/}(1), 184--206.

\bibitem[\protect\citeauthoryear{Drmac and Gugercin}{Drmac and Gugercin}{2016}]{drmac2016new}
Drmac, Z. and S.~Gugercin (2016).
\newblock A new selection operator for the discrete empirical interpolation method---improved a priori error bound and extensions.
\newblock {\em SIAM Journal on Scientific Computing\/}~{\em 38\/}(2), A631--A648.

\bibitem[\protect\citeauthoryear{Goreinov, Oseledets, Savostyanov, Tyrtyshnikov, and Zamarashkin}{Goreinov et~al.}{2010}]{goreinov2010find}
Goreinov, S.~A., I.~V. Oseledets, D.~V. Savostyanov, E.~E. Tyrtyshnikov, and N.~L. Zamarashkin (2010).
\newblock How to find a good submatrix.

\bibitem[\protect\citeauthoryear{Goreinov, Tyrtyshnikov, and Zamarashkin}{Goreinov et~al.}{1997}]{goreinov1997theory}
Goreinov, S.~A., E.~E. Tyrtyshnikov, and N.~L. Zamarashkin (1997).
\newblock A theory of pseudoskeleton approximations.
\newblock {\em Linear algebra and its applications\/}~{\em 261\/}(1-3), 1--21.

\bibitem[\protect\citeauthoryear{Haber, Biton, Rogel, Herbst, Shekhar, Smillie, Burgin, Delorey, Howitt, Katz, et~al.}{Haber et~al.}{2017}]{haber2017single}
Haber, A.~L., M.~Biton, N.~Rogel, R.~H. Herbst, K.~Shekhar, C.~Smillie, G.~Burgin, T.~M. Delorey, M.~R. Howitt, Y.~Katz, et~al. (2017).
\newblock A single-cell survey of the small intestinal epithelium.
\newblock {\em Nature\/}~{\em 551\/}(7680), 333--339.

\bibitem[\protect\citeauthoryear{Han, Wu, and Silke}{Han et~al.}{2020}]{han2020overview}
Han, J., J.~Wu, and J.~Silke (2020).
\newblock An overview of mammalian p38 mitogen-activated protein kinases, central regulators of cell stress and receptor signaling.
\newblock {\em F1000Research\/}~{\em 9}.

\bibitem[\protect\citeauthoryear{Hari-Dass, Shah, Meyer, and Raynes}{Hari-Dass et~al.}{2005}]{hari2005serum}
Hari-Dass, R., C.~Shah, D.~J. Meyer, and J.~G. Raynes (2005).
\newblock Serum amyloid a protein binds to outer membrane protein a of gram-negative bacteria.
\newblock {\em Journal of Biological Chemistry\/}~{\em 280\/}(19), 18562--18567.

\bibitem[\protect\citeauthoryear{Higuera, Gardiner, and Cios}{Higuera et~al.}{2015}]{higuera2015self}
Higuera, C., K.~J. Gardiner, and K.~J. Cios (2015).
\newblock Self-organizing feature maps identify proteins critical to learning in a mouse model of down syndrome.
\newblock {\em PloS one\/}~{\em 10\/}(6), e0129126.

\bibitem[\protect\citeauthoryear{Hooper, Stappenbeck, Hong, and Gordon}{Hooper et~al.}{2003}]{hooper2003angiogenins}
Hooper, L.~V., T.~S. Stappenbeck, C.~V. Hong, and J.~I. Gordon (2003).
\newblock Angiogenins: a new class of microbicidal proteins involved in innate immunity.
\newblock {\em Nature immunology\/}~{\em 4\/}(3), 269--273.

\bibitem[\protect\citeauthoryear{Hotelling}{Hotelling}{1933}]{hotelling1933analysis}
Hotelling, H. (1933).
\newblock Analysis of a complex of statistical variables into principal components.
\newblock {\em Journal of educational psychology\/}~{\em 24\/}(6), 417.

\bibitem[\protect\citeauthoryear{Hsu, Li, and Fuchs}{Hsu et~al.}{2014}]{hsu2014transit}
Hsu, Y.-C., L.~Li, and E.~Fuchs (2014).
\newblock Transit-amplifying cells orchestrate stem cell activity and tissue regeneration.
\newblock {\em Cell\/}~{\em 157\/}(4), 935--949.

\bibitem[\protect\citeauthoryear{Iannello, Crack, De~Haan, and Kola}{Iannello et~al.}{1999}]{iannello1999oxidative}
Iannello, R., P.~J. Crack, J.~B. De~Haan, and I.~Kola (1999).
\newblock {\em Oxidative stress and neural dysfunction in Down syndrome}.
\newblock Springer.

\bibitem[\protect\citeauthoryear{Jolliffe and Cadima}{Jolliffe and Cadima}{2016}]{jolliffe2016principal}
Jolliffe, I.~T. and J.~Cadima (2016).
\newblock Principal component analysis: a review and recent developments.
\newblock {\em Philosophical transactions of the royal society A: Mathematical, Physical and Engineering Sciences\/}~{\em 374\/}(2065), 20150202.

\bibitem[\protect\citeauthoryear{Jones, Townes, Li, and Engelhardt}{Jones et~al.}{2022}]{jones2022contrastive}
Jones, A., F.~W. Townes, D.~Li, and B.~E. Engelhardt (2022).
\newblock Contrastive latent variable modeling with application to case-control sequencing experiments.
\newblock {\em The Annals of Applied Statistics\/}~{\em 16\/}(3), 1268--1291.

\bibitem[\protect\citeauthoryear{Korb and Finkbeiner}{Korb and Finkbeiner}{2011}]{korb2011arc}
Korb, E. and S.~Finkbeiner (2011).
\newblock Arc in synaptic plasticity: from gene to behavior.
\newblock {\em Trends in neurosciences\/}~{\em 34\/}(11), 591--598.

\bibitem[\protect\citeauthoryear{Li, Jones, and Engelhardt}{Li et~al.}{2020}]{li2020probabilistic}
Li, D., A.~Jones, and B.~Engelhardt (2020).
\newblock Probabilistic contrastive principal component analysis.
\newblock {\em arXiv preprint arXiv:2012.07977\/}.

\bibitem[\protect\citeauthoryear{Liu and Chen}{Liu and Chen}{2002}]{liu2002ferredoxin}
Liu, G. and X.~Chen (2002).
\newblock The ferredoxin reductase gene is regulated by the p53 family and sensitizes cells to oxidative stress-induced apoptosis.
\newblock {\em Oncogene\/}~{\em 21\/}(47), 7195--7204.

\bibitem[\protect\citeauthoryear{Liu, Yang, Yu, Dong, Wang, Wang, and Wang}{Liu et~al.}{2023}]{liu2023tau}
Liu, H., Z.~Yang, C.~Yu, H.~Dong, S.~Wang, G.~Wang, and D.~Wang (2023).
\newblock Tau aggravates stress-induced anxiety by inhibiting adult ventral hippocampal neurogenesis in mice.
\newblock {\em Cerebral Cortex\/}~{\em 33\/}(7), 3853--3865.

\bibitem[\protect\citeauthoryear{Liu, Chen, Jia, Qian, Yang, Zhang, Fang, and Liu}{Liu et~al.}{2023}]{liu2023comprehensive}
Liu, Z., S.~Chen, W.~Jia, Y.~Qian, X.~Yang, M.~Zhang, T.~Fang, and H.~Liu (2023).
\newblock Comprehensive analysis reveals ccdc60 as a potential biomarker correlated with prognosis and immune infiltration of head and neck squamous cell carcinoma.
\newblock {\em Frontiers in Oncology\/}~{\em 13}, 1113781.

\bibitem[\protect\citeauthoryear{Loftis and Janowsky}{Loftis and Janowsky}{2003}]{loftis2003n}
Loftis, J.~M. and A.~Janowsky (2003).
\newblock The n-methyl-d-aspartate receptor subunit nr2b: localization, functional properties, regulation, and clinical implications.
\newblock {\em Pharmacology \& therapeutics\/}~{\em 97\/}(1), 55--85.

\bibitem[\protect\citeauthoryear{Logunova, Kapina, Kondratieva, and Apt}{Logunova et~al.}{2023}]{logunova2023h2}
Logunova, N., M.~Kapina, E.~Kondratieva, and A.~Apt (2023).
\newblock The h2-a class ii molecule $\alpha$/$\beta$-chain cis-mismatch severely affects cell surface expression, selection of conventional cd4+ t cells and protection against tb infection.
\newblock {\em Frontiers in Immunology\/}~{\em 14}, 1183614.

\bibitem[\protect\citeauthoryear{Loonen, Stolte, Jaklofsky, Meijerink, Dekker, Van~Baarlen, and Wells}{Loonen et~al.}{2014}]{loonen2014reg3gamma}
Loonen, L.~M., E.~Stolte, M.~T. Jaklofsky, M.~Meijerink, J.~Dekker, P.~Van~Baarlen, and J.~Wells (2014).
\newblock Reg3$\gamma$-deficient mice have altered mucus distribution and increased mucosal inflammatory responses to the microbiota and enteric pathogens in the ileum.
\newblock {\em Mucosal immunology\/}~{\em 7\/}(4), 939--947.

\bibitem[\protect\citeauthoryear{Love, Huber, and Anders}{Love et~al.}{2014}]{love2014moderated}
Love, M.~I., W.~Huber, and S.~Anders (2014).
\newblock Moderated estimation of fold change and dispersion for rna-seq data with deseq2.
\newblock {\em Genome biology\/}~{\em 15}, 1--21.

\bibitem[\protect\citeauthoryear{Luberg, Park, Aleksejeva, and Timmusk}{Luberg et~al.}{2015}]{luberg2015novel}
Luberg, K., R.~Park, E.~Aleksejeva, and T.~Timmusk (2015).
\newblock Novel transcripts reveal a complex structure of the human trka gene and imply the presence of multiple protein isoforms.
\newblock {\em BMC neuroscience\/}~{\em 16}, 1--21.

\bibitem[\protect\citeauthoryear{Mahoney and Drineas}{Mahoney and Drineas}{2009}]{mahoney2009cur}
Mahoney, M.~W. and P.~Drineas (2009).
\newblock Cur matrix decompositions for improved data analysis.
\newblock {\em Proceedings of the National Academy of Sciences\/}~{\em 106\/}(3), 697--702.

\bibitem[\protect\citeauthoryear{McFarland, Paolella, Warren, Geiger-Schuller, Shibue, Rothberg, Kuksenko, Colgan, Jones, Chambers, et~al.}{McFarland et~al.}{2020}]{mcfarland2020multiplexed}
McFarland, J.~M., B.~R. Paolella, A.~Warren, K.~Geiger-Schuller, T.~Shibue, M.~Rothberg, O.~Kuksenko, W.~N. Colgan, A.~Jones, E.~Chambers, et~al. (2020).
\newblock Multiplexed single-cell transcriptional response profiling to define cancer vulnerabilities and therapeutic mechanism of action.
\newblock {\em Nature communications\/}~{\em 11\/}(1), 4296.

\bibitem[\protect\citeauthoryear{McQuail, Beas, Kelly, Simpson, Frazier, Setlow, and Bizon}{McQuail et~al.}{2016}]{mcquail2016nr2a}
McQuail, J.~A., B.~S. Beas, K.~B. Kelly, K.~L. Simpson, C.~J. Frazier, B.~Setlow, and J.~L. Bizon (2016).
\newblock Nr2a-containing nmdars in the prefrontal cortex are required for working memory and associated with age-related cognitive decline.
\newblock {\em Journal of Neuroscience\/}~{\em 36\/}(50), 12537--12548.

\bibitem[\protect\citeauthoryear{Mills, Bartlett, Dissing-Olesen, Wisniewska, Kuznicki, Macvicar, Wang, and Bamji}{Mills et~al.}{2014}]{mills2014cognitive}
Mills, F., T.~E. Bartlett, L.~Dissing-Olesen, M.~B. Wisniewska, J.~Kuznicki, B.~A. Macvicar, Y.~T. Wang, and S.~X. Bamji (2014).
\newblock Cognitive flexibility and long-term depression (ltd) are impaired following $\beta$-catenin stabilization in vivo.
\newblock {\em Proceedings of the National Academy of Sciences\/}~{\em 111\/}(23), 8631--8636.

\bibitem[\protect\citeauthoryear{NCBI}{NCBI}{2024a}]{ncbi_h4c3}
NCBI (2024a).
\newblock {H4C3 gene}.
\newblock \url{https://www.ncbi.nlm.nih.gov/gene/8364}.

\bibitem[\protect\citeauthoryear{NCBI}{NCBI}{2024b}]{ncbi_plat24c}
NCBI (2024b).
\newblock {PLATG4C gene}.
\newblock \url{https://www.ncbi.nlm.nih.gov/gene/8605}.

\bibitem[\protect\citeauthoryear{NCBI}{NCBI}{2024c}]{ncbi_prkcg}
NCBI (2024c).
\newblock {PRKCG gene}.
\newblock \url{https://www.ncbi.nlm.nih.gov/gene/5582}.

\bibitem[\protect\citeauthoryear{Peixoto and Abel}{Peixoto and Abel}{2013}]{peixoto2013role}
Peixoto, L. and T.~Abel (2013).
\newblock The role of histone acetylation in memory formation and cognitive impairments.
\newblock {\em Neuropsychopharmacology\/}~{\em 38\/}(1), 62--76.

\bibitem[\protect\citeauthoryear{Rees, Tandun, Yau, Zachos, and Steiner}{Rees et~al.}{2020}]{rees2020regenerative}
Rees, W.~D., R.~Tandun, E.~Yau, N.~C. Zachos, and T.~S. Steiner (2020).
\newblock Regenerative intestinal stem cells induced by acute and chronic injury: the saving grace of the epithelium?
\newblock {\em Frontiers in Cell and Developmental Biology\/}~{\em 8}, 583919.

\bibitem[\protect\citeauthoryear{Roy, Roy, Rana, Akhter, Hande, and Banerjee}{Roy et~al.}{2018}]{roy2018role}
Roy, S., S.~Roy, A.~Rana, Y.~Akhter, M.~P. Hande, and B.~Banerjee (2018).
\newblock The role of p38 mapk pathway in p53 compromised state and telomere mediated dna damage response.
\newblock {\em Mutation Research/Genetic Toxicology and Environmental Mutagenesis\/}~{\em 836}, 89--97.

\bibitem[\protect\citeauthoryear{Severson, Ghosh, and Ng}{Severson et~al.}{2019}]{severson2019unsupervised}
Severson, K.~A., S.~Ghosh, and K.~Ng (2019).
\newblock Unsupervised learning with contrastive latent variable models.
\newblock In {\em Proceedings of the AAAI Conference on Artificial Intelligence}, Volume~33, pp.\  4862--4869.

\bibitem[\protect\citeauthoryear{Shalin, Zirrgiebel, Honsa, Julien, Miller, Kaplan, and Sweatt}{Shalin et~al.}{2004}]{shalin2004neuronal}
Shalin, S.~C., U.~Zirrgiebel, K.~J. Honsa, J.-P. Julien, F.~D. Miller, D.~R. Kaplan, and J.~D. Sweatt (2004).
\newblock Neuronal mek is important for normal fear conditioning in mice.
\newblock {\em Journal of neuroscience research\/}~{\em 75\/}(6), 760--770.

\bibitem[\protect\citeauthoryear{Shi, Zhu, Qiu, Tan, Wang, Qin, and Dai}{Shi et~al.}{2023}]{shi2023resistin}
Shi, Y., N.~Zhu, Y.~Qiu, J.~Tan, F.~Wang, L.~Qin, and A.~Dai (2023).
\newblock Resistin-like molecules: a marker, mediator and therapeutic target for multiple diseases.
\newblock {\em Cell Communication and Signaling\/}~{\em 21\/}(1), 18.

\bibitem[\protect\citeauthoryear{Snoeck, Goddeeris, and Cox}{Snoeck et~al.}{2005}]{snoeck2005role}
Snoeck, V., B.~Goddeeris, and E.~Cox (2005).
\newblock The role of enterocytes in the intestinal barrier function and antigen uptake.
\newblock {\em Microbes and infection\/}~{\em 7\/}(7-8), 997--1004.

\bibitem[\protect\citeauthoryear{Sorensen and Embree}{Sorensen and Embree}{2016}]{sorensen2016deim}
Sorensen, D.~C. and M.~Embree (2016).
\newblock A deim induced cur factorization.
\newblock {\em SIAM Journal on Scientific Computing\/}~{\em 38\/}(3), A1454--A1482.

\bibitem[\protect\citeauthoryear{Stewart}{Stewart}{1999}]{stewart1999four}
Stewart, G.~W. (1999).
\newblock Four algorithms for the the efficient computation of truncated pivoted qr approximations to a sparse matrix.
\newblock {\em Numerische Mathematik\/}~{\em 83}, 313--323.

\bibitem[\protect\citeauthoryear{Sugino and Sawada}{Sugino and Sawada}{2022}]{sugino2022influence}
Sugino, H. and Y.~Sawada (2022).
\newblock Influence of s100a2 in human diseases.
\newblock {\em Diagnostics\/}~{\em 12\/}(7), 1756.

\bibitem[\protect\citeauthoryear{Suwandi, Galeev, Riedel, Sharma, Seeger, Sterzenbach, Garc{\'\i}a~Pastor, Boyle, Gal-Mor, Hensel, et~al.}{Suwandi et~al.}{2019}]{suwandi2019std}
Suwandi, A., A.~Galeev, R.~Riedel, S.~Sharma, K.~Seeger, T.~Sterzenbach, L.~Garc{\'\i}a~Pastor, E.~C. Boyle, O.~Gal-Mor, M.~Hensel, et~al. (2019).
\newblock Std fimbriae-fucose interaction increases salmonella-induced intestinal inflammation and prolongs colonization.
\newblock {\em PLoS pathogens\/}~{\em 15\/}(7), e1007915.

\bibitem[\protect\citeauthoryear{Suzuki, Tanaka, Poyurovsky, Nagano, Mayama, Ohkubo, Lokshin, Hosokawa, Nakayama, Suzuki, et~al.}{Suzuki et~al.}{2010}]{suzuki2010phosphate}
Suzuki, S., T.~Tanaka, M.~V. Poyurovsky, H.~Nagano, T.~Mayama, S.~Ohkubo, M.~Lokshin, H.~Hosokawa, T.~Nakayama, Y.~Suzuki, et~al. (2010).
\newblock Phosphate-activated glutaminase (gls2), a p53-inducible regulator of glutamine metabolism and reactive oxygen species.
\newblock {\em Proceedings of the National Academy of Sciences\/}~{\em 107\/}(16), 7461--7466.

\bibitem[\protect\citeauthoryear{Vassilev, Vu, Graves, Carvajal, Podlaski, Filipovic, Kong, Kammlott, Lukacs, Klein, et~al.}{Vassilev et~al.}{2004}]{vassilev2004vivo}
Vassilev, L.~T., B.~T. Vu, B.~Graves, D.~Carvajal, F.~Podlaski, Z.~Filipovic, N.~Kong, U.~Kammlott, C.~Lukacs, C.~Klein, et~al. (2004).
\newblock In vivo activation of the p53 pathway by small-molecule antagonists of mdm2.
\newblock {\em Science\/}~{\em 303\/}(5659), 844--848.

\bibitem[\protect\citeauthoryear{Wang, Kang, Kang, and Yang}{Wang et~al.}{2023}]{wang2023s100a6}
Wang, Y., X.~Kang, X.~Kang, and F.~Yang (2023).
\newblock S100a6: molecular function and biomarker role.
\newblock {\em Biomarker Research\/}~{\em 11\/}(1), 78.

\bibitem[\protect\citeauthoryear{Weinberger, Covert, and Lee}{Weinberger et~al.}{2024}]{weinberger2024feature}
Weinberger, E., I.~Covert, and S.-I. Lee (2024).
\newblock Feature selection in the contrastive analysis setting.
\newblock {\em Advances in Neural Information Processing Systems\/}~{\em 36}.

\bibitem[\protect\citeauthoryear{Yasuda, Hayashi, and Hell}{Yasuda et~al.}{2022}]{yasuda2022camkii}
Yasuda, R., Y.~Hayashi, and J.~W. Hell (2022).
\newblock Camkii: a central molecular organizer of synaptic plasticity, learning and memory.
\newblock {\em Nature Reviews Neuroscience\/}~{\em 23\/}(11), 666--682.

\bibitem[\protect\citeauthoryear{Yeo, Itahana, Guo, Han, Iwamoto, Nguyen, Bao, Kleiber, Wu, Bay, et~al.}{Yeo et~al.}{2016}]{yeo2016transglutaminase}
Yeo, S.~Y., Y.~Itahana, A.~K. Guo, R.~Han, K.~Iwamoto, H.~T. Nguyen, Y.~Bao, K.~Kleiber, Y.~J. Wu, B.~H. Bay, et~al. (2016).
\newblock Transglutaminase 2 contributes to a tp53-induced autophagy program to prevent oncogenic transformation.
\newblock {\em Elife\/}~{\em 5}, e07101.

\bibitem[\protect\citeauthoryear{Yin, Godbout, and Sheridan}{Yin et~al.}{2024}]{yin2024interleukin}
Yin, W., J.~P. Godbout, and J.~F. Sheridan (2024).
\newblock Interleukin-1 beta in psychosocial stress.
\newblock In {\em Stress: Immunology and Inflammation}, pp.\  53--63. Elsevier.

\bibitem[\protect\citeauthoryear{Yu, Xiong, and Zhang}{Yu et~al.}{2023}]{yu2023role}
Yu, H., M.~Xiong, and Z.~Zhang (2023).
\newblock The role of glycogen synthase kinase 3 beta in neurodegenerative diseases.
\newblock {\em Frontiers in Molecular Neuroscience\/}~{\em 16}, 1209703.

\end{thebibliography}

\newpage

\appendix

\section{Data and Code Availability}
\label{appendix:code}
Code and Data availability can be found at \\ \href{https://anonymous.4open.science/r/CCUR-E5CD/README.md}{https://anonymous.4open.science/r/CCUR-E5CD/README.md}.

\section{CUR Algorithm}\label{apdx:alg}

\Cref{appendix:cur} presents the detailed CUR algorithm.
\begin{algorithm}
\caption{CUR}\label{appendix:cur}
\begin{algorithmic}[1]
\State \textbf{Input} data: $\{X_i\}_{i=1}^n$, number of singular vectors $k$, number of columns selected $c$, number of rows to select $r$.
\State Compute SVD of $X = U\Sigma V^T$.
\State Compute leverage scores for each column in $X$
$$l_d = \sum_{\xi = 1}^k (v^{\xi})_d^2 , \quad d = 1 \ldots p
$$
\State Select highest $c$ leverage scores to construct $C$.
\State Repeat steps 2 and 3 for $X^T$ and construct $R$ with the highest $r$ leverage scores.
\State Compute $U = C^+ X R^+$ where $A^+$ denotes a Moore-Penrose generalized inverse of the matrix $A$.
\State Return $C, U, R$.
\end{algorithmic}
\label{alg:cur}
\end{algorithm}

\section{Hyperparameter Tuning}
\label{appendix:hyper}
There are 3 main hyperparameters that we must consider when running CCUR: $k$, the number of singular vectors, $\epsilon$, the upweighting constant, $c$, the number of genes selected, and $r$, the number of rows selected. For each of our experiments, $k = 7, \epsilon = 10^{-6}, c = 10$. The number of rows selected in Mouse Protein was 25. For Small Molecules and Pathogen, $r = 200$. $r$ was chosen based on the size of the datasets. Since the foreground in Mouse Protein contained significantly less observations, $r$ was set to a smaller number.

The parameter $k$ determines how many singular vectors are used to compute the leverage scores. In high-dimensional datasets, such as those encountered in scRNA-seq, smaller values of $k$ are often sufficient to capture most of the variance in the data. For our experiments, we found that $k$ in the range of 5 to 10 was effective, with minimal variation in results when $k$ was adjusted within this range. 

The upweighting constant $\epsilon$ is added to the denominator of the leverage score ratio to prevent inflated ratios when the background leverage scores are near zero. While $\epsilon$ typically has a negligible effect on the results when background scores are nonzero, it becomes critical in cases where some background scores are close to or exactly zero. Without this adjustment, ratios can become excessively large, even when the foreground scores are also small, leading to the selection of genes that may not be biologically meaningful. In practice, we observed that $\epsilon = 10^{-6}$ provides a robust safeguard against these extreme cases without significantly altering results in most datasets. Users can inspect the distribution of background leverage scores before setting $\epsilon$. If the distribution includes values close to zero, retaining or slightly increasing $\epsilon$ is advisable. If all background scores are comfortably above zero, $\epsilon$ could be reduced or even omitted.

The parameter $c$ controls how many genes are ultimately selected by CCUR. For our experiments and simulations, we chose $c = 10$ as it provided a manageable set of genes for downstream analysis while still capturing key biological insights. Users can adjust $c$ depending on their specific goals. A smaller $c$ might be preferable for focused studies that aim to highlight a few highly relevant genes, while a larger $c$ could be used for broader exploratory analysis. However, users should consider that increasing $c$ may also include genes with weaker contrastive signals, potentially diluting the interpretability of the results. Similarly, for selecting 
$r$, users can adjust this parameter based on the specific needs of their study.

\section{Additional Simulation and Experimental Details}
\label{appendix:experiments}
All experiments were conducted on a MacBook Pro equipped with an Apple M4 Pro chip (14-core CPU with 10 performance and 4 efficiency cores), 24 GB of unified memory, and an integrated Apple M4 Pro GPU supporting Metal 3. The system featured a built-in 3024 × 1964 Liquid Retina XDR display and macOS running OS Loader version 11881.81.4. For the simulation setup, the number of singular vectors was set to 10, the latent dimension was set to 5, the foreground and background sample size was set to 500 each, and the number of features was set to 100. For sample selection, the number of columns selected in CCUR first was 10. Additionally, in all experiments where CPCA was used, $\alpha$ was set to 1. To select genes using CPCA, we looked at the top genes with the highest coefficient in absolute value. A summary of the datasets used is shown in \cref{tab:summary_datasets}.

\begin{table}[ht]
\centering
\renewcommand{\arraystretch}{1.5} 
\setlength{\tabcolsep}{10pt}       
\begin{tabular}{@{}lcccc@{}}       
\toprule
\textbf{Experiment} & \( n \) & \( m \) & \( p \) & \(sg\) \\
\midrule
Mouse Protein (\ref{sec:mp}) & 270 & 135 & 77 & 2 \\
Small Molecules (\ref{sec:sm}) & 3096 & 2831 & 1000 & 2 \\
Pathogen (\ref{sec:pathogen}) & 4481 & 3240 & 1000 & 2 \\
\bottomrule
\end{tabular}
\vspace{3pt}
\caption{Summary of characteristics of datasets for each application. \(n\) and \(m\) represent the sizes of the foreground and background datasets, respectively, \(p\) is the number of features (genes), and \(sg\) is the number of subgroups in the foreground.}
\label{tab:summary_datasets}
\end{table}

\subsubsection*{Mouse Protein}
\label{appendix:mp}

\begin{figure}[H]
    \centering
    \fcolorbox{white}{white}{
        \adjustbox{bgcolor=white}{\includegraphics[width=\linewidth]{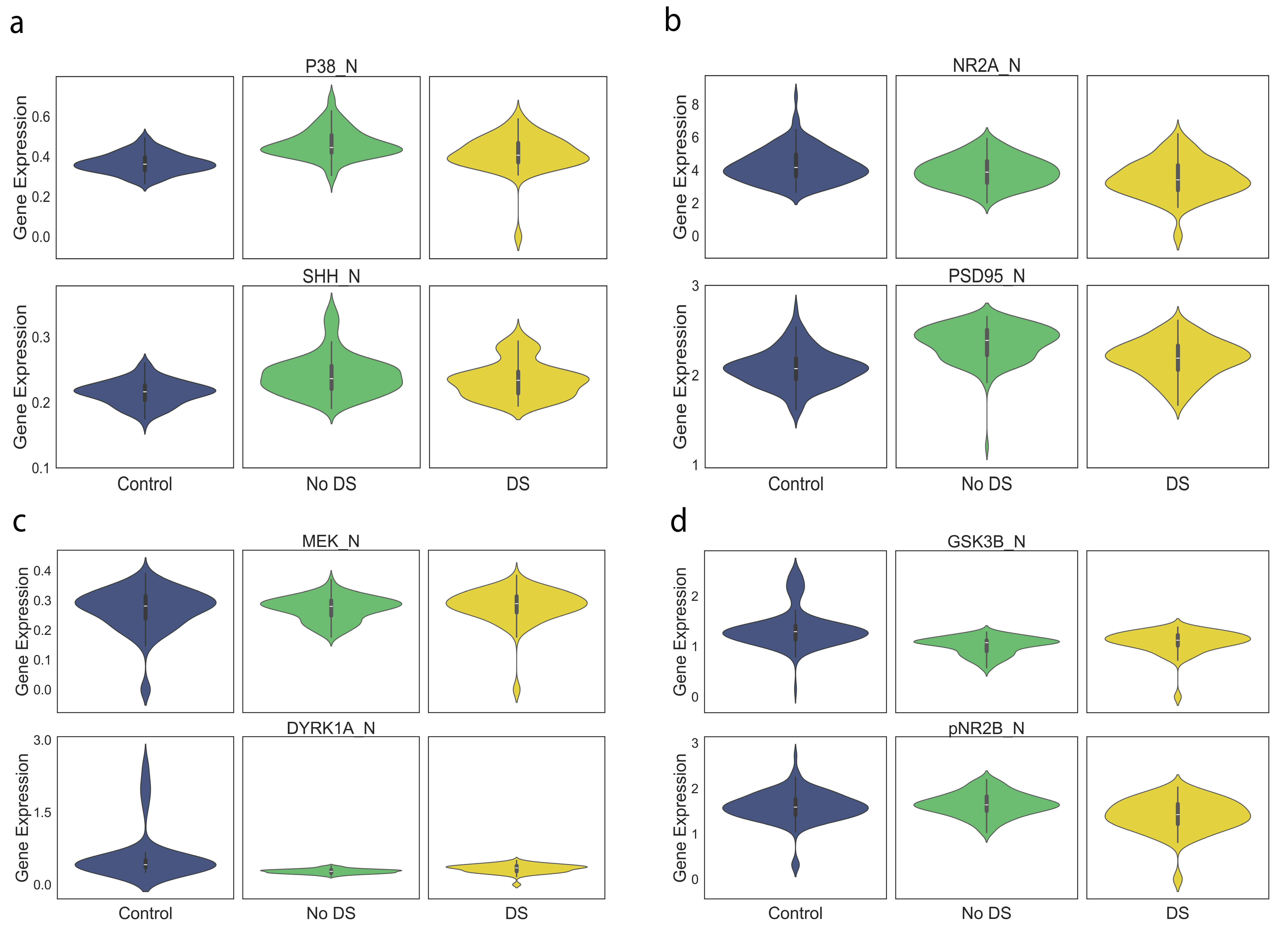}}
    }
    \caption{Mouse Protein Dataset. Gene expressions of additional genes selected by (a) CCUR (b) CUR (c) CFS (d) CPCA. A comparison of the control and the two subgroups (no DS and DS) in the foreground is shown.}
    \label{fig:mp_suppl}
\end{figure}

In this section, we include additional experimental details and results from the mouse protein dataset. As previously highlighted, CCUR consistently selects genes with strong associations to stress responses and neurological processes. For instance, \textit{p38} is a key regulator of cellular stress responses, including inflammation, DNA damage repair, and metabolic stress \citep{han2020overview}. Similarly, \textit{Shh} has emerged as a critical modulator in neurodegenerative diseases and plays an essential role in maintaining neural system processes \citep{chen2018emerging}. Additionally, \textit{Arc} is fundamental for memory consolidation and is intricately linked to neuronal activity \citep{korb2011arc}. These findings demonstrate that the genes identified by CCUR are highly relevant to stress responses, brain function, and neurodegenerative conditions such as Down Syndrome, underscoring the method's ability to uncover biologically significant pathways. The expressions of these genes are displayed in \cref{fig:mp_suppl}.

\subsubsection*{Small Molecules}
\label{appendix:sm}
\begin{figure}[H]
    \centering
\includegraphics[width=\linewidth]{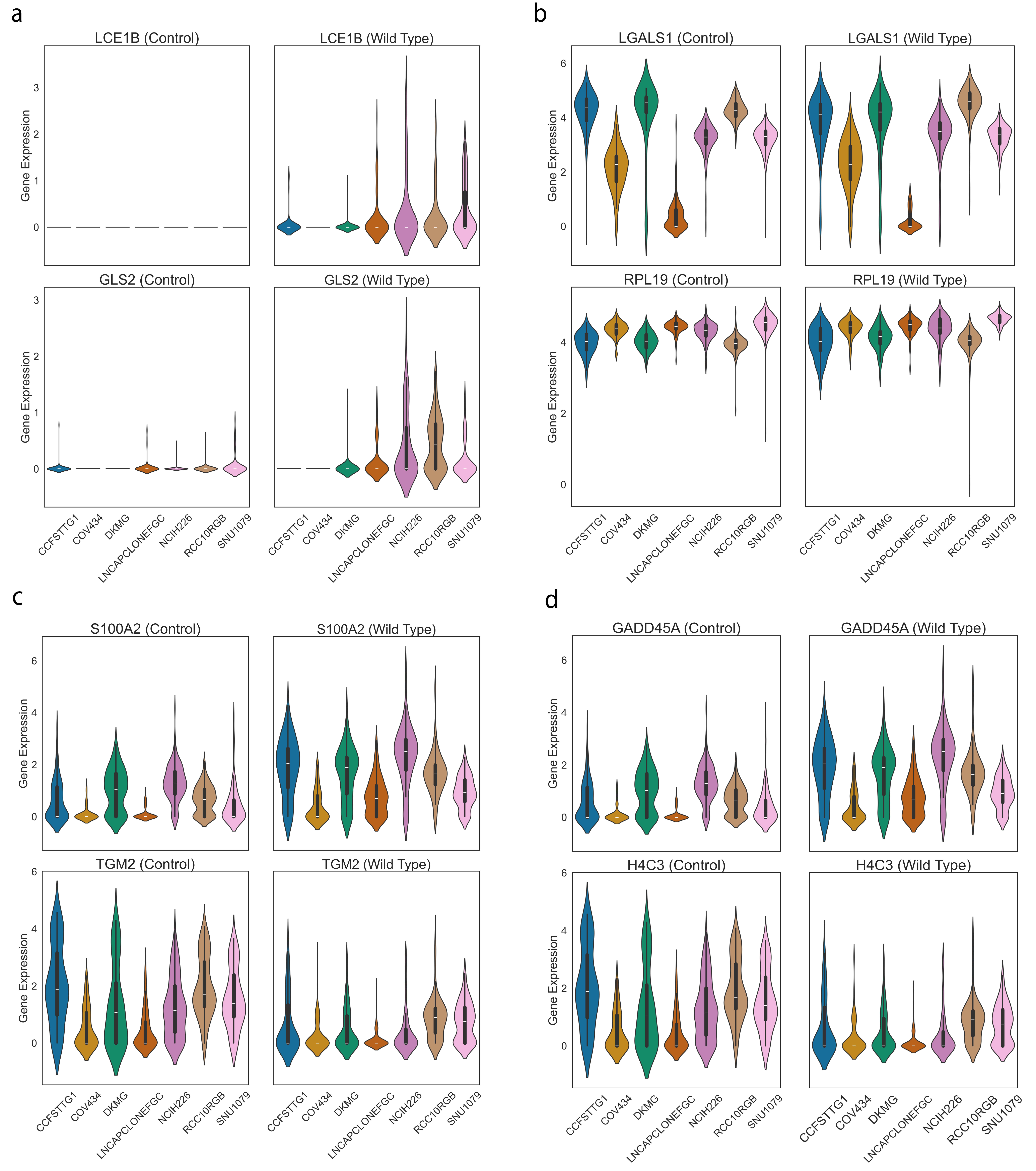}
    \caption{Small Molecules Dataset. Gene expressions of additional genes selected by (a) CCUR (b) CUR (c) CFS (d) CPCA. \textit{S100A2} was also selected by CUR. \textit{Fdxr} in \cref{fig:sm} was also selected by CPCA. A comparison of the control and the WT cells in the foreground is shown.}
    \label{fig:sm_suppl}
\end{figure}

In this section, we provide further analysis on the genes selected by CCUR, CUR, CPCA, and CFS. \textit{Lce1} genes and \textit{Gls2} have been identified as downstream targets of p53 \citep{deng2014late, suzuki2010phosphate}. Further investigation revealed a specific p53-binding site within the \textit{Lce1} gene cluster that functions as an enhancer, confirming that p53 directly transactivates \textit{Lce1} genes. Additional results are presented in \cref{fig:sm_suppl}.

\subsubsection*{Pathogen}
\label{appendix:pathogen}
\begin{figure}[h!]
    \centering
\includegraphics[width=\linewidth]{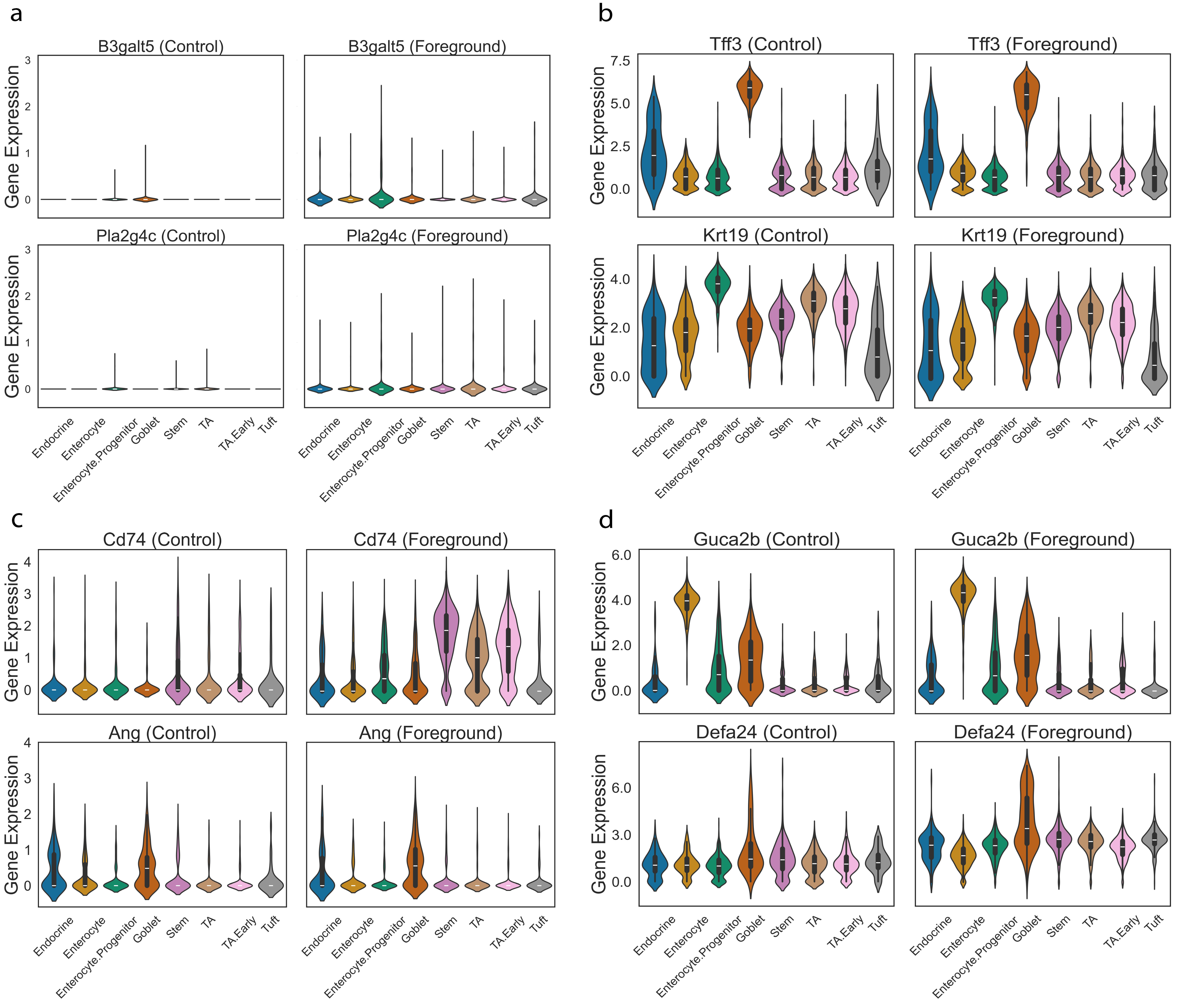}
    \caption{Pathogen Dataset. Gene expressions of additional genes selected by (a) CCUR (b) CUR (c) CFS (d) CPCA. \textit{Guca2b} and \textit{Defa24} were also selected by CUR. A comparison between the control and foreground across each cell is shown.}
    \label{fig:pathogen_suppl}
\end{figure}

In this section, we present additional experimental plots highlighting the genes selected by each method (\cref{fig:pathogen_suppl}). Both \textit{B3galt5} is crucial for enhancing mucosal defense and modulating immune responses during infections, including Salmonella \citep{suwandi2019std}. The \textit{Pla2g4c} gene produces an enzyme that helps create molecules involved in inflammation and immune responses, suggesting it may play a role in fighting infections like Salmonella \citep{ncbi_plat24c}. Notably, we observe that the differences in expression between the foreground and background are more pronounced in specific cell types, suggesting that these gene-induced changes may be cell-specific.

\end{document}